\documentclass[prl,aps,twocolumn,floatfix,superscriptaddress,showpacs]{revtex4}
\usepackage{revsymb,amssymb,amscd}
\usepackage{graphicx,epsf,epsfig,natbib}
\usepackage{bm}

\begin{document}

\title{Spin diffusion of correlated two-spin states in a dielectric crystal}
\author{G. S. Boutis}
\affiliation{Department of Nuclear Engineering, Massachusetts Institute of Technology, Cambridge, MA 02139 }
\author{D. Greenbaum}
\affiliation{Department of Physics, Massachusetts Institute of Technology, Cambridge, MA 02139}
\author{H. Cho}
\affiliation{Department of Nuclear Engineering, Massachusetts Institute of Technology, Cambridge, MA 02139 }
\author{D. G. Cory}
\affiliation{Department of Nuclear Engineering, Massachusetts Institute of Technology, Cambridge, MA 02139 }
\author{C. Ramanathan}
\email[Corresponding author.  E-mail address: ]{sekhar@mit.edu}
\affiliation{Department of Nuclear Engineering, Massachusetts Institute of Technology, Cambridge, MA 02139 }

\begin{abstract}

Reciprocal space measurements of spin diffusion in a single crystal of calcium fluoride (CaF$_2$) have been extended to dipolar ordered states.  The experimental results for the component of the spin diffusion  parallel to the external field are $D_{D}^{||}=29 \pm 3 \times 10^{-12}$
cm$^{2}$/s for the [001] direction and $D_{D}^{||}=33 \pm 4 \times
10^{-12}$ cm$^{2}$/s for the [111] direction. The measured diffusion rates for dipolar order are
 faster than those for Zeeman order and are considerably faster than predicted by simple theoretical models. It is suggested that constructive interference in the transport of the two spin state is responsible for this enhancement.  As expected the anisotropy in the diffusion rates is observed to be significantly less for dipolar order compared to the Zeeman case.
\end{abstract}
\maketitle

In this work we measure and compare the spin diffusion rates for dipolar and Zeeman energy in a dielectric crystal.  Spin diffusion provides a well-posed problem in the study of multi-body dynamics, as the Hamiltonian of the system is well known.   The study of the diffusion of the Zeeman and dipolar ordered states is essentially the study of the evolution of different initial states under the secular dipolar Hamiltonian. In particular the Zeeman ordered state consists of single spin population terms only, while the dipolar ordered state consists of correlated two spin states \cite{Cho}. In a strong magnetic field these quantities are independently conserved, and have different diffusion rates and spin-lattice relaxation times.  The spins are in dynamical equilibrium under the action of the secular dipolar Hamiltonian, and the constants of the motion are only defined for the total spin system.  If the identity of individual spins can be distinguished, the spins are clearly seen to evolve in time as in the case where we write a position dependent phase on the spins.  This is the essence of our spin diffusion experiments.    Previous attempts at measuring spin diffusion of Zeeman \cite{Leppelmeier,Gates} and dipolar energy \cite{Furman} relied on theoretical models that related the diffusion rate to observed relaxation times.   Zhang and Cory used reciprocal space techniques to perform the first direct experimental measurement of spin diffusion in a crystal of CaF$_2$ \cite{Zhang}. 

\begin{figure}
\epsfxsize=15pc
\epsfbox{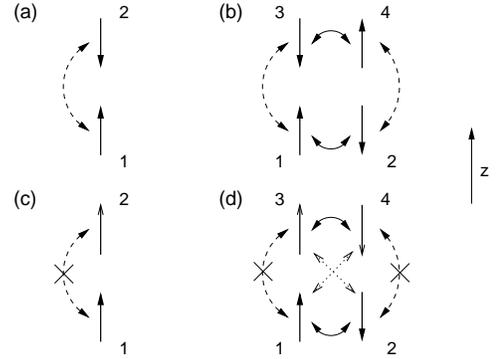}
\caption{(a) For the diffusion of Zeeman order along the z direction spins 1 and 2 need to undergo a flip-flop.  (b) For the diffusion of dipolar order along the z direction, there are two possibilities, both spins 1 and 3, and spins 2 and 4 can flip, or spins 1 and 2, and spins 3 and 4 can flip.  There are thus two different paths to the same final state and interference effects may be observed. (c) If spins 1 and 2 are initially in the same state (both $\uparrow$ or both $\downarrow$),  no evolution takes place. (d) Even if states 3 and 4 are initially  $\uparrow$ and $\downarrow$, two different evolution paths are present.  Spins 1 and 2 can flip and spins 3 and 4 can flip or spins 1 and 4 and spins 2 and 3 can flip.  Thus dipolar diffusion dynamics are less easily quenched. }
\end{figure}

There have been a number of attempts to calculate the rate of spin diffusion in single
crystals, for both the Zeeman energy and the dipolar energy of the spin system using theoretical models \cite{Bloem,Khutsishvili-1,deGennes,Redfield,Lowe,Walgraef1,Yu,Walgraef2}, and classical simulation \cite{Tang,SodicksonII}.  These calculations suggest that the spin diffusion rates of Zeeman and dipolar order should approximately be the same.  However, an examination of the physical processes leading to the diffusion shows that the dynamics can be quite different.
Figure 1 shows a simple illustrative model that suggests that the diffusion of dipolar order should be faster (and more complicated) than that of Zeeman order, due to the increase in the number of possible paths for the propogation of the dipolar ordered state.  While a faster diffusion rate would seem to make the measurement of dipolar diffusion easier, the measurement is more challenging as the experiment has to be carried out in a time comparable to $T_{\mathit{1D}}$ (usually an order of magnitude smaller than $T_{1}$).  

The basic reciprocal space NMR spin diffusion experiments for Zeeman and dipolar order are outlined below.   The Hamiltonian of the system is given by $\mathcal{H} = \mathcal{H}_{Z} + \mathcal{H}_{D}$, where $\mathcal{H}_{Z}  =  \hbar \omega \sum_i I_{z}^i$ is the Zeeman interaction, and 
\begin{eqnarray}
\mathcal{H}_{D} & = &  \hbar \sum_{i< j} b_{ij} \left( 2I_{z}^{i}I_{z}^{j} - \frac{1}{2} ( I_{+}^{i}I_{-}^{j}+I_{-}^{i}I_{+}^{j}) \right) 
\end{eqnarray}
is the dipolar interaction, $\omega$ is the Larmor frequency of the spins and $b_{ij} = \frac{\gamma^2 \hbar}{2}\frac{1 -3 \cos^2 \theta_{ij}}{r_{ij}^3}$ is the strength of the dipolar coupling between spins $i$ and $j$.   In the high field ($|\mathcal{H}_{Z}| >> |\mathcal{H}_{D}|$), high temperature ($\beta\hbar\omega << 1$) limit, the initial equilibrium density matrix is given by
\begin{equation}
\rho_{0} \approx \frac{1}{\mathcal{Z}}\left(\mathbf{1} - \beta \hbar \omega \sum_{i} I_{z}^{i} \right) = \frac{\mathbf{1}}{\mathcal{Z}} - \delta\rho
\end{equation}
where $\mathcal{Z}$ is the partition function and $\mathbf{1}$ is the identity. In the following discussion we follow the evolution of $\delta\rho$.

After the $(\pi/2)_y$ pulse, the magnetization grating is written onto the spin system by interleaving strong magnetic field gradient pulses in the long delay times within a magic echo train \cite{Magic_echo,Boutis}.  The magic echo train suspends evolution of the dipolar coupling, and the effective Hamiltonian of the system  is just the gradient term,
\begin{equation}
\mathcal{H}_\nabla =\gamma\hbar\frac{\partial B_{z}}{\partial z} \sum_{i} I_{z}^{i} z_{i}
\end{equation}
where $\gamma$ is the nuclear gyromagnetic ratio, $\partial B_{z}/\partial z$ is the gradient field strength, and $z_{i}$ is the spatial location of spin $i$.  This creates spatially modulated Zeeman magnetization.  If the total time the gradient is applied is $\delta$, the wave number corresponding to the applied gradient is $k = \gamma \delta \left(\partial B_{z}/\partial z \right)$.  
Following the application of the magnetic field gradient, the state is
\begin{equation}
\delta\rho = \frac{\beta\hbar\omega}{\mathcal{Z}} \sum_i I_+^i e^{i\phi_i}
\end{equation}
where $\phi_{i}=kz_{i}$ tags the spatial location of spin $i$.  The state in Eq.\ (4) is no longer a constant of the motion under a collective measurement of the spins.  A $(\pi/2)_{\bar{y}}$ pulse flips one component of the transverse magnetization back along the longitudinal ($z$) axis.  The remaining transverse component decays rapidly yielding
\begin{equation}
\delta\rho = \frac{\beta\hbar\omega}{\mathcal{Z}} \sum_i I_z^i \cos\phi_i
\end{equation}
These steps are common to both experiments. 

{\em Zeeman diffusion}

We allow the above state to evolve under $\mathcal{H}_{D}$ for a time $\Delta$.   Following spin diffusion, we can express the state as
\begin{equation}
\delta\rho(\Delta) = \frac{\beta\hbar\omega}{\mathcal{Z}} \sum_{i,j} c_{ij}(\Delta) I_z^j \cos\phi_i
\end{equation}
where the coefficients $c_{ij}(\Delta)$ represent the relative amount of polarization initially at site $i$ that has been transported to site $j$ after time $\Delta$.  We have also neglected higher order spin terms in the above equation, as these do not contribute when we trace over the spins.  In the absence of the phase tag, these terms would sum to zero, as the Zeeman state is a constant of the motion.

In the final step, a $(\pi/2)_y$ pulse is applied followed by a magnetic field gradient, equal in amplitude but reversed in direction to the encoding step. This gradient labels the new location of the polarization, in analogy with a scattering experiment.  Following a $(\pi/2)_{\bar{y}}$ pulse, the longitudinal component of the magnetization is 
\begin{equation}
\delta\rho(\Delta) = \frac{\beta\hbar\omega}{\mathcal{Z}} \sum_{i,j} c_{ij}(\Delta) I_z^j \cos\phi_i\cos\phi_j
\end{equation}
In the absence of spin diffusion, $c_{ij}= \delta_{ij}$, the phase terms would be identical and the original state would be recovered (scaled by 1/2).  In the presence of spin diffusion the residual phase $\cos(\phi_i-\phi_j)$ reflects the spatial transport of the magnetization.  In the experiment, the total $z$ magnetization is measured, tracing over the spatial coordinates of the sample.  Under diffusive dynamics the observed attenuation is $\exp(-k^{2}D\Delta)$.

{\em Dipolar diffusion experiment}

The dipolar diffusion experiment requires two additional steps.  The modulated Zeeman magnetization (Eq.\ (5)) is first converted to encoded dipolar order by applying either the two-pulse Jeener-Broekaert (JB) sequence \cite{jeener} or an adiabatic demagnetization in the rotating frame (ADRF)~\cite{AndersonHartmann}.   Following the creation of dipolar order, the state of the system is
\begin{equation}
\delta\rho = \frac{\beta\hbar}{\mathcal{Z}} \sum_{i<j}b_{ij}\left(2I_z^iI_z^j-\frac{1}{2}(I_+^iI_-^j + I_-^iI_+^j)\right) \cos\phi_i
\end{equation}
This evolves under $\mathcal{H}_{D}$ for time $\Delta$ during which spin diffusion takes place. The state can then be expressed as
\begin{eqnarray}
\delta\rho & = & \frac{\beta\hbar}{\mathcal{Z}} \sum_{i<j;k<l} b_{ij} d_{ijkl}(\Delta) \cos\phi_i \;\; \times \nonumber \\ & & \hspace*{0.5in}\left(2I_z^kI_z^l-\frac{1}{2}(I_+^kI_-^l + I_-^kI_+^l)\right)
\end{eqnarray}
We have again left out higher order terms that do not contribute to the measured magnetization.  These terms would sum to zero in the absence of a phase tag, as the dipolar ordered state is a constant of the motion.
 The two-spin dipolar ordered state is converted back to a single-spin Zeeman state using either a $\pi/4$ pulse or an adiabatic remagnetization in the rotating frame (ARRF).  The $\pi/4$ pulse only converts a portion of the dipolar ordered state back to Zeeman order, while the ARRF allows a complete refocusing to the Zeeman state.  The gradient modulation is subsequently unwound  yielding
\begin{equation}
\delta \rho = \frac{\beta \hbar}{\mathcal{Z}} \sum_{i<j;k<l} b_{ij} d_{ijkl}(\Delta) \epsilon'_{kl}\left(I_z^k \cos \phi_k  + I_z^l \cos \phi_l \right) \cos \phi_i
\end{equation}
where $\epsilon_{kl}$ is a dimensionless quantity proportional to $b_{kl}$.
In the limit of infinitesimal polarization the dipolar ordered states created are local two-spin correlations.  The length scale of these correlations (\AA)  is much smaller than the length scale of the spatial modulation (100's of nm), yielding $\phi_{k}=\phi_{l}$.  As a consequence  the phase spread due to the creation and refocusing of dipolar order can also be safely ignored in the experiments here.  In the absence of spin diffusion, $\phi_{i}=\phi_{k}$ and the original state is recovered (scaled by 1/2).  As in the Zeeman case, the residual phase term $\cos (\phi_{i} - \phi_{k})$ encodes the spatial transport in the presence of spin diffusion.

\begin{figure}
\epsfxsize=20pc
\epsfbox{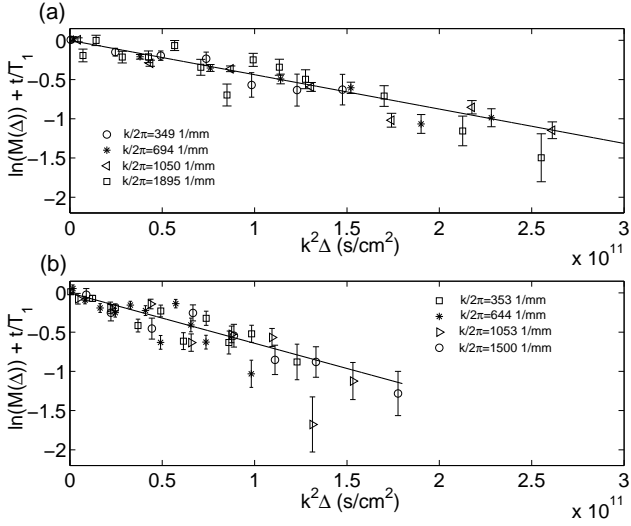}
\caption{(a) Experimental results of the spin diffusion
rate of Zeeman order with the crystal aligned along the [111]
direction. The pulsed gradient fields generated are $\circ$ 2886
G/cm, $\ast$ 5661 G/cm, $\triangleleft$ 8510 G/cm, $\square$ 8510
G/cm.  (b) Experimental results of the spin diffusion
rate of Zeeman order with the crystal aligned along the [001]
direction. The pulsed gradient fields generated are $\square$ 2886 G/cm,
$\ast$ 5661 G/cm,  $\triangleright$ 8695 G/cm, $\circ$ 7474 G/cm. The error in $k$ is approximately 3 \%. }
\end{figure}

We performed the measurements at room temperature at 4.7 T ($^{19}$F 188.35 MHz) using a Bruker Avance spectrometer, and a home-built z-gradient probe. The gradient coil had an inductance of 26 $\mu$H, a resistance of 1.6~$\Omega$, and an efficiency of 370 $\pm$ 10.5 G/cmA \cite{wurongsprobe}.   The solenoidal RF coil geometry used has been shown to produce a homogeneous field over a small sample volume ~\cite{Prigl}.  The experiments were performed on a 1 mm$^3$ single crystal of CaF$_{2}$ having $T_{1} =  256.2\pm3.4$ s along the [001] direction and $T_{1} =  287.5\pm7$ s along the [111] direction.   The measured $T_{\mathit{1D}}$ of our crystal was $9.4\pm0.4$ s along the [001] direction and $9.3\pm0.1$ s along the [111] direction.  The $\pi/2$ pulse was 1.65 $\mu$s, the spacing between magic echoes was 60 $\mu$s and the length of each gradient pulse was 35 $\mu$s. The coil constant was determined by calibration with a sample of water whose molecular diffusion coefficient is well known.  A phase alternating scheme was applied to correct for RF pulse errors in the magic-echo sequences \cite{Boutis}.    A $\pi/4$ pulsed refocusing of the dipolar ordered state was used in the JB experiment and an ARRF was used with the ADRF.

\begin{figure}
\epsfxsize=20pc
\epsfbox{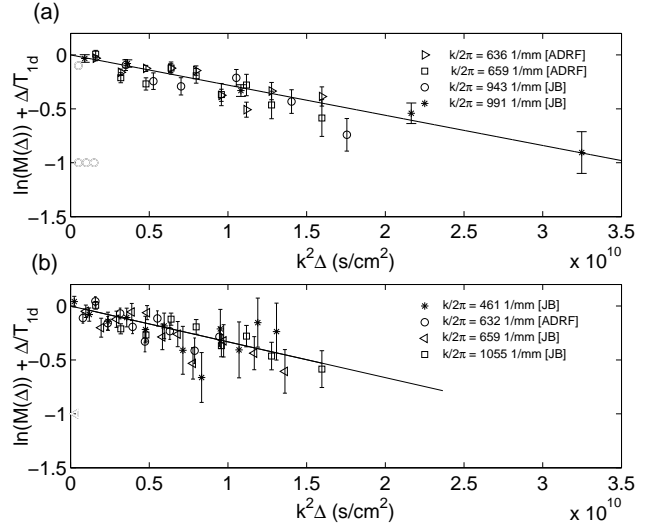}
\caption{(a) Experimental results of the spin diffusion
rate of dipolar order with the crystal aligned along the [001]
direction. The pulsed gradient fields generated are  $\triangleright$ 5698 G/cm,
$\square$ 5698 G/cm, $\circ$ 8473 G/cm , $\ast$ 8399 G/cm. (b) Experimental results of the spin
diffusion rate of dipolar order with the crystal aligned along the
[111] direction.  The pulsed gradient fields generated are  $\ast$ 2886 G/cm,
$\circ$ 8436 G/cm, $\triangleleft$ 5661 G/cm, $\square$ 8362 G/cm.  The error in $k$ is approximately 3 \%.}
\end{figure}

\begin{table}
\caption{Summary of the experimental results of the spin diffusion
rate of spin-spin energy, $D_{D}$, and Zeeman energy, $D_{Z}$ for
single crystal calcium fluoride.}
\begin{tabular} {||c|c|c|c||} \hline This
measurement&[001]&[111]&$D_{001}/D_{111}$  \\ \hline
\hspace*{0.1in} $D^{||}_{D}$ \hspace*{0.1in} ($ \times 10^{-12}$cm$^{2}$/s) & 29 $\pm$ 3& 33 $\pm$ 4 & 0.88 $\pm$ 0.14 \\
\hspace*{0.1in} $D^{||}_{Z}$ \hspace*{0.1in} ($ \times 10^{-12}$cm$^{2}$/s) & 6.4 $\pm$ 0.9 & 4.4 $\pm$ 0.5 & 1.45 $\pm$ 0.26  \\
 \hline

Prior measurement \cite{Zhang}& [001] &[111]&$D_{001}/D_{111}$ \\
\hline
\hspace*{0.1in} $D^{||}_{Z}$ \hspace*{0.1in}  ($ \times 10^{-12}$cm$^{2}$/s) & 7.1 $\pm$ 0.5 & 5.3 $\pm$ 0.3 &1.34 $\pm$ 0.12
\\
\hline

Theoretical studies of $D_{Z}^{||}$& [001] & [111] &$D_{001}/D_{111}$ \\ \hline

Ref. \cite{Walgraef1} ($ \times 10^{-12}$cm$^{2}$/s) & 6.98 &  4.98 & 1.4\\
Ref. \cite{Yu}  ($\times 10^{-12}$cm$^{2}$/s) & 8.22  & 6.71 & 1.22 \\
Ref. \cite{Tang}  ($\times 10^{-12}$cm$^{2}$/s) & 7.43& -- & -- \\ \hline

Theoretical studies of $D_{D}^{||}$& [001] & [111]& $D_{001}/D_{111}$ \\ \hline

Ref. \cite{Walgraef2}  ($\times 10^{-12}$cm$^{2}$/s) & 8.53 &  8.73 & 0.98 \\
Ref. \cite{Tang}   ($\times 10^{-12}$cm$^{2}$/s) & 13.3& -- & -- \\ \hline

Ratio of $D_{D}$ to $D_{Z}$& [001] & [111] & \\ \hline
This measurement & $4.5\pm0.8$ & $7.5\pm1.3$& -- \\
Ref. \cite{Walgraef2}& 1.22 & 1.75 & --  \\
Ref. \cite{Tang} & 1.79 & -- & -- \\ \hline
\end{tabular}
\label{tableofresults}
\end{table}

The measured signal at long times, $\Delta \gg (2\pi/b_{ij}^{max})$, is
\begin{equation}
S = S_{0} \exp(-k^{2}D\Delta) \exp(-\Delta/T_{1x})
\end{equation}
as the system also undergoes spin-lattice relaxation during the time $\Delta$ ($T_{1x} = T_{1}$ in the Zeeman experiment and $T_{1x} = T_{1D}$ in the dipolar experiment).  It is necessary to correct the signal amplitude for this attenuation in order to obtain the correct diffusion coefficient. The crystal used in this experiment has a much longer $T_{1}$ than the crystal used previously for the Zeeman diffusion measurement \cite{Zhang} ($T_{1} = 114.2\pm5.3$ s along the [001] direction and $T_{1}= 156.8\pm9.7$ s along the [111] direction),  since the dipolar diffusion experiment needs to be carried out in a time comparable to $T_{\mathit{1D}}$ ($T_{\mathit{1D}}<T_{1}$).  In figures 2 and 3 we plot $\ln(S) + \Delta/T_{1x}$ versus $k^{2}\Delta$ for the Zeeman and dipolar experiments when the external magnetic field is applied along the [001] and [111] crystal orientations.  The diffusive character of the dynamics is shown by the universal scaling observed with the different gratings.

Table 1 summarizes the experimental results and shows previous theoretical and numerical results for comparison.  The measurements and theory agree reasonably well for Zeeman order, both in the absolute value and the orientational dependence of the spin diffusion rate.  Though the various theories predict the ratio of spin diffusion rates for dipolar and Zeeman order should be between one and two,  the measurement yields a ratio larger than four.  In contrast to Zeeman diffusion, dipolar diffusion exhibits a much smaller anisotropy.   This is expected as the anisotropy of the field generated by a single magnetic dipole is larger than that created by a pair of dipoles.

In the description of the Zeeman and dipolar diffusion experiments above, the details of the dynamics have
been aggregated and included in the coefficients $c_{ij}$ and $d_{ijkl}$.  Evaluation of these terms would entail solving the many body problem for spin diffusion.  However, in order to compare the two diffusion processes, it is illustrative to evaluate these coefficients in the short time limit.  Evaluating the second commutator for the persistence of the initial states, we obtain 
\begin{equation}
c_{ii}(t)  =  1 - \frac{t^2}{4}\sum_{k \neq i}(b_{ik})^2 + O(t^4)
\end{equation}
\begin{equation}
d_{ijij}(t)  \approx  1 - \frac{t^2}{4} \sum_{k \neq i,j}
(b_{ik}^2 + b_{jk}^2 + b_{ik}b_{jk}) + O(t^4).
\end{equation}
The equation for $d_{ijij}$ above is not exact, as there are also some
two spin terms present that are not proportional to the dipolar ordered 
state.  If the dipolar ordered state had been proportional to
$I_{z}^{i}I_{z}^{j}$, then we would have $d_{ijkl} = c_{ik}c_{jl}$.
The coefficient $d$ in Eq.\ (13) differs from a simple
product of the $c$'s at this order by the cross-terms
$b_{ik}b_{jk}$.  These cross terms could either increase or decrease 
the decay rate  as the dipolar couplings can be both positive and negative.  
Thus both constructive and destructive interference effects could be obtained in 
principle.  This suggests that the rapid diffusion rate observed for dipolar order is a 
consequence of a constructive interference effect that enhances the diffusion rate over 
that of unentangled spin pairs.  
Currently available calculations do not take this interference effect into account properly.  
For instance, the moment method \cite{Yu} introduces an ad-hoc cutoff function in frequency space 
to force an exponential solution, 
even though the coherent signal evolution should actually be an even function of time.
Indeed, the cross terms in Eq.\ (13) cancel out the level of the second moment, and it is
unclear whether they remain in the higher moments, as the calculation of higher order
moments has historically proved challenging.  Incorporating such interference effects in the calculation
is important in our system where the dynamics remain coherent and reversible over the 
timescale of the experiment -- about 10$^6$ times the correlation time of the dipolar coulping.

In conclusion, we have performed a direct measurement of spin diffusion of dipolar energy in a single crystal of CaF$_2$.  The rate is observed to be significantly faster than expected from previous theoretical studies, though the reduced anisotropy is well predicted.   Given the highly mixed nature of the initial density matrix in these experiments and the large Hilbert space, it is difficult to characterize the dynamics in more detail.   Measuring spin diffusion as a function of increasing polarization would allow us to follow the spin dynamics as they are constrained to progressively smaller subspaces, enabling a more careful exploration of the Hilbert space dynamics.

\vspace*{0.04in}
{\noindent \bf Acknowledgements}

\noindent We thank N. Boulant, P. Cappellaro, J. V. Emerson, E. M. Fortunato, M. Kindermann, M. A. Pravia, G. Teklemariam, W. Zhang and  Professors J. S. Waugh and L. S. Levitov for interesting discussions, and the NSF and DARPA for financial support.

\end{document}